\documentclass[a4paper,12pt]{article}
\newcommand{\newsection}[1]{
\vspace{5mm}
\pagebreak[3]
\refstepcounter{section}
\setcounter{subsection}{0}
\begin{flushleft}
{\large\bf \thesection. #1}
\end{flushleft}
\nopagebreak
\medskip
\nopagebreak}

\newcommand{\newsubsection}[1]{
\vspace{5mm}
\pagebreak[3]
\addtocounter{subsection}{1}
\addcontentsline{toc}{subsection}{\protect
\numberline{\arabic{section}.\arabic{subsection}}{#1}}
\noindent{\em 
\thesubsection. #1}
\nopagebreak
\vspace{2mm}
\nopagebreak}
\newcommand{\ud}{\mathrm{d}}

\makeatletter

\def\clap#1{\hbox to 0pt{\hss #1\hss}}%

\def\ligne#1{%
  \hbox to \hsize{%
    \vbox{\centering #1}}}%

\def\haut#1#2#3{%
  \hbox to \hsize{%
    \rlap{\vtop{\raggedright #1}}%
    \hss
    \clap{\vtop{\centering #2}}%
    \hss
    \llap{\vtop{\raggedleft #3}}}}%
\def\maketitle{%
  \thispagestyle{empty}\vbox to \vsize{%
    \@preprint
    \vfill
    \ligne{\Large \@title}
    \vspace{5mm}
    \ligne{\@author}
    \vspace{1mm}\ligne{\texttt{<\@email>}}
    \vspace{5mm}
    \ligne{\@blurb}
    \vspace{1cm}
    \@abstract
    \vfill
    }%
  }

\def\date#1{\def\@date{#1}}
\def\author#1{\def\@author{#1}}
\def\title#1{\def\@title{#1}}
\def\location#1{\def\@location{#1}}
\def\blurb#1{\def\@blurb{#1}}
\def\email#1{\def\@email{#1}}
\def\abstract#1{\def\@abstract{#1}}
\def\preprint#1{\def\@preprint{#1}}

\date{\today}
\author{}
\title{}
\location{Durham}
\blurb{}
\email{no email address}
\makeatother
  \title{A Charged Doubly Spinning Black Ring}
  \author{James Hoskisson}
  \email{James.Hoskisson@durham.ac.uk}
  \date{21st August 2008}
  \preprint{\begin{flushright}
  DCPT-08/49
  \end{flushright}}
  \location{Durham}
  \blurb{
    Department of Mathematical Sciences \\
    University of Durham \\
    Science Laboratories \\
    South Road \\
    Durham \\
    DH1 3LE
    }%
  \abstract{
  \begin{center}
	\bf{Abstract}
	\end{center}
	\vspace{3mm}
This paper calculates the general form of a 5d metric when fundamental string and momentum charges are added. This is accomplished using the standard method of boosting and T-dualising a solution to Einstein's equations, where the solution has three Killing vectors and is expressed in a generic form. The thermodynamical properties of the charged solution are derived and the physical implications of the solution are then examined with the two-charge dual spinning black ring being used as an example.}

\usepackage{graphicx}
\usepackage[font={footnotesize,sf},labelfont={bf,rm}]{caption}

\begin{document}

\maketitle
\setcounter{footnote}{0}
\newsection{Introduction}

The development of string theory, along with its implication that there exist more than three spatial dimensions, has led researchers to investigate solutions to general relativity in more than four dimensions. The majority of the effort has been focussed on finding solutions in five dimensions, since this is the next simplest scenario, and several black hole \cite{Myers:1986un,Emparan:2001wn,Pomeransky:2006bd,Morisawa:2007di} and multi-black hole \cite{Elvang:2007rd,Iguchi:2007is,Elvang:2007hs,Izumi:2007qx} solutions have been discovered. These solutions can be characterised as having either $S^3$ topology or $S^2\times S^1$ topology (or a combination of the two in the case of the multi-black hole solutions). The discovery of a black hole in five dimensions with a non-spherical topology was remarkable because in four dimensions all of the black hole solutions have a topology of $S^2$ and are described in terms of the mass, angular momentum, and charge. A comprehensive review of all the higher dimensional black hole solutions is given in \cite{Emparan:2008eg}.

Asymptotically flat black hole solutions are of interest since there is a microscopic understanding of the origin of their entropy. Strominger and Vafa were the first to investigate this in \cite{Strominger:1996sh}, where they examined a class of supersymmetric spherical 5d black holes with non-zero charge and showed that it was possible to derive their entropy by counting the degeneracy of BPS states. Further work building on this has been done for black holes with both $S^3$ and $S^2\times S^1$ topology in papers such as \cite{Mateos:2001qs,Elvang:2003mj,Horowitz:2007xq,Larsen:2005qr,Elvang:2007hs,Reall:2007jv}. These papers examine a mixture of charged and neutral black hole solutions with the charged solutions calculated for each specific metric. The procedure presented in this paper gives a five parameter metric which can be used to analyse the microscopic origin of the entropy for a much broader range of black hole solutions, since the only constraint on the uncharged starting metric is that it has three Killing vectors.

Although the method in this paper has already been used to find charged versions of the Myers-Perry and the singly spinning black ring, it hasn't yet been used to generate charged solutions for the doubly spinning black ring and other recently discovered multi-black hole solutions. Given that the procedure for adding string charges to a neutral metric is virtually identical for all asymptotically flat metrics it seems useful to give the solution in as general a form as possible. The first part of this paper gives the necessary theory for adding charges to a generic metric with three Killing vectors before going on to apply the results to the specific case of the doubly spinning ring.

The doubly spinning ring was chosen because it is the next simplest black hole solution for which a charged version has not been calculated and its properties examined. The singly spinning ring with an event horizon of $S^2\times S^1$ was discovered by Emparan and Reall in 2001 \cite{Emparan:2001wn}, with the rotation being in the $S^1$ direction. Indeed, in order for it to remain stable it had to rotate at a specific speed determined by the geometry of the ring. Unfortunately, their solution didn't allow for the ring to rotate in the $S^2$ direction which one would assume the most general solution should. A black ring solution which did allow for the ring to rotate freely in both the $S^2$ and $S^1$ directions was discovered in 2006 by Pomeransky and Sen'kov \cite{Pomeransky:2006bd}.

Despite being algebraically more complicated than the singly spinning ring, Pomeransky and Sen'kov were able to present the solution in a reasonably compact form, thus allowing other authors to examine the physical properties of the ring, such as in \cite{Elvang:2007hs} and \cite{Emparan:2008eg}. The solution derived by Pomeransky and Sen'kov was only specified for the balanced dual rotating ring where the angular parameters are chosen such that the ring is balanced. The unbalanced solution for doubly rotating rings had already been calculated in \cite{Figueras:2005zp}, which concurs with the respective limit of a more general solution given in \cite{Morisawa:2007di}. The solution of Morisawa et al. permits both balanced and unbalanced rings and allows a more thorough analysis than for the Pomeransky Sen'kov black ring as it is possible to look at the limits of the doubly spinning ring where it is inherently unbalanced, as well as the limit where the ring "collapses" to the spherical Myers Perry black hole.

Although a two charge solution hasn't previously been calculated for the doubly spinning ring using the method presented here, a three charge solution to minimal supergravity has been calculated in \cite{Bouchareb:2007ax}. Their method produces a three charge version of the doubly spinning black ring with all of the charges being equal. This differs from the results presented in this paper, as the method presented here allows the two charges to be set independently of one another. It is possible in principle to add a third charge through a further series of dualisations and boosts but the unavoidable by-product of adding the extra charge is the introduction of Dirac-Meisner string singularities (as is seen in \cite{Bouchareb:2007ax}), so the three charge metric is not considered in this paper.

The first two sections of this paper are concerned with developing the procedure whereby string and momentum charges are added to a generic five dimensional metric. The basic idea is to lift the neutral metric to ten dimensions by adding five extra flat dimensions, and then applying a series of boosts and T-duality transformations to the metric. The ten dimensional metric is then Kaluza-Klein reduced back down to five dimensions with the boosts in the extra spatial dimensions appearing as fundamental string and momentum charges. The resulting charged metric is then presented, along with a derivation of the physical properties of this generic charged solution.

The third section is concerned with the Pomeransky Sen'kov dual spinning ring solution. The dual spinning black ring metric is presented along with some explanation of the adapted coordinates which are used. A brief recap of some of the relevant physical properties is also given\footnote{A more detailed analysis of the physical properties of the dual rotating black ring solution can be found in \cite{Elvang:2007hs}.} before presenting the two charge dual spinning ring metric and its associated fields.

The final section of this paper looks at the physical properties of the two charge generic metric derived in section \ref{Chargemetsec}, using the dual rotating black ring as a specific example. The charged solution has two extra parameters relating to the two string charges, so the physical properties of the charged metric are compared with those of the neutral starting metric for different values of these charges. The differences between the generic charged metric and the neutral metric are independent of the form of the original metric coefficients so the analysis can be applied to any metric that is charged up in the manner described in section \ref{Theory}.

\newsection{Theory of Generating Charges}
\label{Theory}

Having obtained a solution to Einstein's field equations, it is possible to generate a supersymmetric solution that has different charges corresponding to the different string sectors. The method presented here, which adds charges to a generic uncharged metric, is essentially the same as that presented in \cite{Breckenridge:1996sn,Cvetic:1996xz} and more pertinently applied to the singly spinning ring in \cite{Elvang:2003mj}. The main idea is to lift the five dimensional metric to ten dimensions through the inclusion of five flat dimensions, which will be labelled as \{w,6,7,8,9\}. The $w$ dimension is singled out as this plays an important part in constructing the charges. The remaining four dimensions are compactified on a $T^4$ and play a passive role in the generation of the various charges. Having constructed a ten dimensional metric, the application of a series of boosts and duality transformations will then produce a new solution to the string equations of motion. Once this is done, the ten dimensional solution can then be Kaluza-Klein compactified to reduce it back down to the desired five dimensional metric with various string charges.

Any solution to Einstein's equations will automatically satisfy the equations of motion for low energy superstrings when lifted to ten dimensions, as long as the gauge fields are turned off. This can easily be seen by examining the action for the low energy NS-NS superstring, when compactified on $T^4$ \cite{Elvang:2003mj}
\begin{equation}
S_6=\frac{1}{2\kappa_6^2}\int{\ud^6 x\sqrt{-g^{(6)}}e^{-2\Phi}\left(R^{(6)}+4(\nabla\Phi)^2-\frac{1}{12}{H^{(6)}}^2\right)}
\label{stringaction}
\end{equation}
where $\Phi$ is the scalar dilaton, $\kappa_6$ is related to the six dimensional Planck length, and $H^{(6)}$ is a 3-form flux given by $H^{(6)}=\ud B^{(6)}$, where $B^{(6)}$ and $F^{(6)}$ are 2-form fields that couple electrically to the dual rotating ring. The four extra flat dimensions have been suppressed here as they don't enter into any of the subsequent calculations. If $\Phi$ and $B$ are set to zero then the action given by (\ref{stringaction}) reduces to the Einstein-Hilbert action for a six dimensional metric.

The charges are produced by Lorentz boosting the six dimensional metric in the $w$ direction to produce a black-string or black-tube, depending on the topology of the starting solution. This is achieved via a coordinate transformation where $\alpha_1$ is the boost parameter. This gives the solution linear momentum in the $w$ direction but in order to create a non-trivial charge a subsequent T-dualisation must be performed. The application of the T-duality transformation converts the type IIA solution\footnote{This is not a unique choice since the starting solution can equally be thought of as a type IIB solution. In which case, the T-duality transformation would produce a type IIA solution with fundamental charge.} with linear momentum, to a type IIB solution with a fundamental string charge. The T-duality transformations are given by \cite{Horowitz:1992jp}
\begin{equation}
\begin{array}{ll}
g_{ww}\rightarrow\frac{1}{g_{ww}} & \hspace{2cm} g_{w\alpha}\rightarrow\frac{B_{w\alpha}}{g_{ww}} \\
g_{\alpha\beta} \rightarrow g_{\alpha\beta}-\left(\frac{g_{w\alpha}g_{w\beta}-B_{w\alpha}B_{w\beta}}{g_{ww}}\right) & \hspace{2cm} B_{w\alpha}\rightarrow \frac{g_{w\alpha}}{g_{ww}} \\
B_{\alpha\beta}\rightarrow B_{\alpha\beta} - 2\frac{g_{w[\alpha}B_{\beta]w}}{g_{ww}} & \hspace{2cm} \Phi\rightarrow \Phi-\frac{1}{2}\log{g_{ww}} \\
\end{array}
\end{equation}

Having T-dualised the metric, a second charge can be added by boosting again in the $w$ direction, with parameter $\alpha_2$. This gives a black tube with linear momentum in the $w$ direction and a fundamental charge $F(w)$. To obtain the five dimensional solution with two charges it is then simply a case of KK reducing this 10d metric along the $w$ direction.

In order to carry out the KK compactification, assume that the $w$ direction forms a circle of radius $R_w$. The Kaluza-Klein ansatz is then given by
\begin{equation}
g^{(6)}_{MN}\ud x^M\ud x^N=g_{\mu\nu}\ud x^\mu\ud x^\nu + e^{2\sigma}(\ud z+A_\mu^{(1)}\ud x^\mu)^2
\label{KKansatz}
\end{equation}
where the Greek indices cover \{$t$,$\rho$,$z$,$\psi$,$\phi$\} and the Latin indices cover \{$x^\mu$,$w$\}. Here, $e^{2\sigma}=g_{ww}$ and $A^{(1)}$ is a 1-form field, induced by the compactification of the six dimensional metric, which sources the $P(w)$ charge.

Applying (\ref{KKansatz}) to the 6d action (\ref{stringaction}), gives
\begin{equation}
S_5=\frac{1}{2\kappa_5^2}\int{\ud^5 x\sqrt{-g}e^{-2\Phi+\sigma}\left(R^{(5)}+4(\nabla\Phi)^2-4\nabla\Phi\nabla\sigma-\frac{1}{12}H^2-\frac{1}{4}e^{2\sigma}(F^{(1)})^2-\frac{1}{4}e^{-2\sigma}(F^{(2)})^2\right)}
\label{KKaction}
\end{equation}
where $\kappa_5^2=\kappa_6^2/(2\pi R_w)$ and the $^{(6)}$ superscript has been dropped where it is obvious that the fields are now five dimensional.

The reduction of the string action to five dimensions has created two new fields $F^{(1)}$ and $F^{(2)}$ where $F_{\mu\nu}^{(1)}=\partial_\mu A_\nu^{(1)}-\partial_\nu A_\mu^{(1)}$ and $F_{\mu\nu}^{(2)}=\partial_\mu A_\nu^{(2)}-\partial_\nu A_\mu^{(2)}$, with $A_\mu^{(2)}=B_{\mu w}^{(6)}$. The three form $H$ has now picked up a Chern-Simons term, so is now given by $H=\ud B - A^{(1)}\wedge F^{(2)}$.

It is possible to transform the string action into something more like the Einstein action by defining an effective dilaton $\Phi_{eff}=\Phi-\sigma/2$, thus allowing the string metric to be transformed into the Einstein frame via $g_{\mu\nu}^E=e^{-\frac{4}{3}\Phi_{eff}}g_{\mu\nu}$. This means that the action in the Einstein frame becomes
\begin{equation}
S_5=\frac{1}{2\kappa_5^2}\int{\ud^5 x\sqrt{-g}\left(R^{(5)}+4(\nabla\Phi)^2-4\nabla\Phi\nabla\sigma-\frac{1}{12}H^2-\frac{1}{4}e^{2\sigma}(F^{(1)})^2-\frac{1}{4}e^{-2\sigma}(F^{(2)})^2\right)}
\end{equation}
Note that all of the fields inside of the brackets have also changed with the change of frame.

\newsection{Charging Up A Three Killing Field Metric}
\label{Chargemetsec}
\newsubsection{The [F(w),P] Charged Metric}

Having established the formalism required to charge up a metric which solves Einstein's field equations, it is now possible to apply it to a general metric with three Killing vectors. Any metric with three Killing vectors given by $\partial_t$, $\partial_\psi$, and $\partial_\phi$ can be written in the form
\begin{equation}
ds^2=g_{tt}\ud t^2+2g_{t\phi}\ud t\ud\phi+g_{\phi\phi}\ud \phi^2 + 2g_{t\psi}\ud t\ud \psi + g_{\psi\psi}\ud\psi^2 + 2g_{\psi\phi}\ud\psi\ud\phi + g_{\rho\rho}\ud \rho^2 + g_{zz}\ud z^2
\label{genmet}
\end{equation}
where all of the metric functions $g_{\mu\nu}$ are solely functions of $\rho$ and $z$. The non-Killing directions $\rho$ and $z$ are inspired by the canonical coordinates used in the inverse scattering technique, although the form given in (\ref{genmet}) differs in that $g_{\rho\rho}$ and $g_{zz}$ are not necessarily equal, allowing any three Killing vector solution to be used.

It is now possible to charge up this general metric, using the technique of boosting and T-dualising described in the previous section, to give a solution to type IIB string theory with fundamental and momentum charges in the $w$ direction. Having done this, the new 6d charged metric is given by
\begin{eqnarray}
ds^2 & = & \frac{1}{g_{ww}\cosh^2{\alpha_1}+g_{tt}\sinh^2{\alpha_1}}\left[(g_{ww}g_{tt}\cosh^2{\alpha_2}+\sinh^2{\alpha_2})\ud\tilde{t}^2+2g_{t\psi}g_{ww}\cosh{\alpha_1}\cosh{\alpha_2}\ud \tilde t\ud\tilde\psi + \right. \nonumber \\
&& 2g_{t\phi}g_{ww}\cosh{\alpha_1}\cosh{\alpha_2}\ud \tilde t\ud\tilde\phi + (g_{ww}g_{\psi\psi}\cosh^2{\alpha_1} + (g_{tt}g_{\psi\psi}-g_{t\psi}^2)\sinh^2{\alpha_1})\ud{\tilde\psi}^2 + \nonumber \\
&&  2(g_{tt}g_{ww}+1)\cosh{\alpha_2}\sinh{\alpha_2}\ud \tilde t\ud\tilde w + 2(g_{ww}g_{\psi\phi}\cosh^2{\alpha_1} + (g_{tt}g_{\psi\phi}-g_{t\psi}g_{t\phi})\sinh^2{\alpha_1})\ud {\tilde\psi}\ud {\tilde\phi} + \nonumber \\
&& 2g_{t\psi}g_{ww}\cosh{\alpha_1}\sinh{\alpha_2}\ud {\tilde\psi}\ud{\tilde w} + (g_{ww}g_{\phi\phi}\cosh^2{\alpha_1} + (g_{tt}g_{\phi\phi}-g_{t\phi}^2)\sinh^2{\alpha_1})\ud{\tilde\phi}^2 + \nonumber \\
&& 2g_{t\phi}g_{ww}\cosh{\alpha_1}\sinh{\alpha_2}\ud \tilde \phi\ud \tilde w + \left. (\cosh^2{\alpha_2}+g_{ww}g_{tt}\sinh^2{\alpha_2})\ud \tilde w^2 \right]+g_{\rho\rho}\ud{\tilde \rho}^2+g_{zz}\ud{\tilde z}^2
\label{chgmet}
\end{eqnarray}
with the auxiliary two form field given by
\begin{eqnarray}
B_{\tilde t\tilde \psi} & = & \frac{g_{t\psi}\sinh{\alpha_1}\sinh{\alpha_2}}{g_{ww}\cosh^2{\alpha_1}+g_{tt}\sinh^2{\alpha_1}} \\
B_{\tilde t\tilde \phi} & = & \frac{g_{t\phi}\sinh{\alpha_1}\sinh{\alpha_2}}{g_{ww}\cosh^2{\alpha_1}+g_{tt}\sinh^2{\alpha_1}} \\
B_{\tilde w\tilde t} & = & \frac{(g_{ww}+g_{tt})\sinh{\alpha_1}\cosh{\alpha_1}}{g_{ww}\cosh^2{\alpha_1}+g_{tt}\sinh^2{\alpha_1}} \\
B_{\tilde w\tilde \psi} & = & \frac{g_{t\psi}\sinh{\alpha_1}\cosh{\alpha_2}}{g_{ww}\cosh^2{\alpha_1}+g_{tt}\sinh^2{\alpha_1}} \\
B_{\tilde w\tilde\phi} & = & \frac{g_{t\phi}\sinh{\alpha_1}\cosh{\alpha_2}}{g_{ww}\cosh^2{\alpha_1}+g_{tt}\sinh^2{\alpha_1}}
\end{eqnarray}
and the scalar dilaton given by
\begin{equation}
e^{-2\tilde\Phi}=g_{ww}\cosh^2{\alpha_1}+g_{tt}\sinh^2{\alpha_1}
\label{dilaton}
\end{equation}

The metric given in (\ref{chgmet}) only has a fundamental charge in the $w$ direction, so to create the P($w$) charge the metric has to be Kaluza-Klein reduced back down to five dimensions. In practice, the supplementary $w$ dimension is always added to the metric by setting $g_{ww}=1$ so, for the sake of simplicity, this constraint has been applied in all of the following equations. Bearing this in mind, the compactified [F($w$),P($w$)] charged metric in the string frame is given by
\begin{eqnarray}
ds_5^2 &= & \frac{1}{h_1h_2}\left\{h_2(g_{tt}\cosh^2{\alpha_2}+\sinh^2{\alpha_2})\ud\tilde{t}^2+2h_2\cosh{\alpha_1}\cosh{\alpha_2}(g_{t\psi}\ud \tilde t\ud\tilde\psi + g_{t\phi}\ud \tilde t\ud\tilde\phi) \right. \nonumber \\
&& \hspace{1.2cm} \left. -\sinh^2{\alpha_2}[(g_{tt}+1)\cosh{\alpha_2}\ud \tilde t+g_{t\psi}\cosh{\alpha_1}\ud \tilde\psi+g_{t\phi}\cosh{\alpha_1}\ud \tilde\phi]^2 \right. \nonumber \\
&& \hspace{1.2cm} \left. + E_{\psi\psi}\ud{\tilde\psi}^2 + 2E_{\psi\phi}\ud {\tilde\psi}\ud {\tilde\phi} + E_{\phi\phi}\ud{\tilde\phi}^2 \right\} + g_{\rho\rho}\ud \rho^2 + g_{zz}\ud z^2
\label{KKmetric}
\end{eqnarray}
where
\begin{eqnarray}
h_n & = &\cosh^2{\alpha_n}+g_{tt}\sinh^2{\alpha_n} \\
E_{\mu\nu} & = & h_2\left[g_{\mu\nu}\cosh^2{\alpha_1}+(g_{tt}g_{\mu\nu}-g_{t\mu}g_{t\nu})\sinh^2{\alpha_1}\right]
\end{eqnarray}
and $\mu\in\{\psi,\phi\}$.

The compactification of the six dimensional metric has introduced two new 1-form fields $A^{(1)}$ and $A^{(2)}$, as well as the 2-form field $B$, and the scalar field $\Phi$. These 1-form fields are given by
\begin{eqnarray}
A^{(1)} & = & \frac{1}{h_1}\Bigl[(1+g_{tt})\sinh{\alpha_1}\cosh{\alpha_1}\ud t + g_{t\psi}\sinh{\alpha_1}\cosh{\alpha_2}\ud \psi + g_{t\phi}\sinh{\alpha_1}\cosh{\alpha_2}\ud \phi\Bigr] \\
A^{(2)} & = & \frac{1}{h_2}\Bigl[(1+g_{tt})\sinh{\alpha_2}\cosh{\alpha_2}\ud t + g_{t\psi}\sinh{\alpha_2}\cosh{\alpha_1}\ud\psi + g_{t\phi}\sinh{\alpha_2}\cosh{\alpha_1}\ud\phi\Bigr]
\end{eqnarray}
with the two form $B$ being reduced to
\begin{equation}
B_{\tilde t\tilde\mu} = \frac{g_{t\mu}\sinh{\alpha_1}\sinh{\alpha_2}}{h_1}
\end{equation}

The dilaton is unchanged by the compactification process so with $g_{ww}=1$, it is now given by
\begin{equation}
e^{-2\tilde\Phi}=\cosh^2{\alpha_1}+g_{tt}\sinh^2{\alpha_1}=h_1
\label{compdilaton}
\end{equation}
and the other scalar field, introduced by the compactification, is given by
\begin{equation}
e^{2\sigma}=\frac{\cosh^2{\alpha_2}+g_{tt}\sinh^2{\alpha_2}}{\cosh^2{\alpha_1}+g_{tt}\sinh^2{\alpha_1}}=\frac{h_2}{h_1}
\label{compscalar}
\end{equation}

\newsubsection{Physical Properties of The Charged Metric}

The process of charging up the metric only affects the metric coefficients involving $t$, $\psi$, and $\phi$, so any of the properties of the metric that depend upon the coefficients involving $\rho$ and $z$ are unchanged. In most cases this means that the position of the event horizon is unchanged, since the coordinates of the neutral metric are usually chosen so that the event horizon is described by a hypersurface where one of the non-Killing directions is held constant. This is exemplified by the dual rotating ring, described in the next section. Having said this, the addition of string charges to the neutral metric does alter the thermodynamic properties of the metric.

If the mass, angular momenta, and area of the neutral metric are given by $M$, $J_{\psi}$, $J_{\phi}$, and $A$ respectively, then it is possible to calculate how these will change with the addition of extra charges. It is assumed in the following that the metric given in (\ref{genmet}) is asymptotically flat, which in turn implies that the charged metric (\ref{chgmet}) is also asymptotically flat.

For an asymptotically flat metric, the ADM mass can be derived by examining the $g_{tt}$ coefficient near asymptotic infinity. This function will then fall off as
\begin{equation}
g_{tt}=-1+\frac{8GM}{3\pi r^2}+{\cal O} \left(\frac{1}{r^4}\right)
\label{gtt}
\end{equation}
at infinity, so the Taylor expansion of the metric function can then be compared to this and the mass $M$ extracted. Since the charged metric is also asymptotically flat, its mass can be calculated in a similar manner. Expressing the $g_{\tilde t\tilde t}$ coefficient of the charged metric in terms of the original metric gives
\begin{equation}
g_{\tilde t\tilde t} = \frac{g_{tt}}{h_1h_2}
\label{gtteqn}
\end{equation}
If it is assumed that $g_{tt}$ takes the form given by (\ref{gtt}), then the above equation becomes
\begin{equation}
g_{\tilde t\tilde t} = -1+\frac{4GM(\cosh{2\alpha_1}+\cosh{2\alpha_2})}{3\pi r^2}+...
\end{equation}
Comparing this with (\ref{gtt}), the charged metric mass $\tilde M$ can be defined as
\begin{equation}
\tilde M = \frac{M}{2}(\cosh{2\alpha_1}+\cosh{2\alpha_2})
\label{ADMM}
\end{equation}

The angular momenta in the $\psi$ and $\phi$ directions can be calculated using a similar process, but this time comparing the different coefficients for $g_{t\psi}$ and $g_{t\phi}$ respectively. The necessary expressions for the charged metric coefficients are given by
\begin{eqnarray}
g_{\tilde t\tilde \phi} & = & \frac{g_{t\phi}\cosh{\alpha_1}\cosh{\alpha_2}}{h_1h_2}
\label{gtphieqn} \\
g_{\tilde t\tilde \psi} & = & \frac{g_{t\psi}\cosh{\alpha_1}\cosh{\alpha_2}}{h_1h_2}
\label{gtpsieqn}
\end{eqnarray}
In this case, the form of the $g_{t\psi}$ and $g_{t\phi}$ coefficients at infinity is
\begin{equation}
g_{t\mu}\sim\frac{J_{\mu}}{r^2}+{\cal O}\left(\frac{1}{r^4}\right)
\label{gtang}
\end{equation}
where $\mu\in\{\psi,\phi\}$. Bearing this in mind, the charged metric coefficients after substituting for $g_{tt}$ from (\ref{gtt}) become
\begin{eqnarray}
g_{\tilde t\tilde \phi} & = & \frac{J_{\phi}}{r^2}\cosh{\alpha_1}\cosh\alpha_2+... \\
g_{\tilde t\tilde \psi} & = & \frac{J_{\psi}}{r^2}\cosh{\alpha_1}\cosh\alpha_2+...
\end{eqnarray}
These equations can then be compared with (\ref{gtang}) to construct expressions for the charged metric angular momenta
\begin{eqnarray}
\tilde J_\phi & = & J_{\phi}\cosh{\alpha_1}\cosh\alpha_2 \\
\tilde J_\psi & = & J_{\psi}\cosh{\alpha_1}\cosh\alpha_2
\label{ADMJ}
\end{eqnarray}

Unfortunately, the above method cannot be used to calculate how the area varies when string charges are added to the neutral metric, since the area is given by
\begin{equation}
A=\int{\sqrt{|\gamma|}}=\int{\sqrt{g_{zz}\left[g_{\psi\psi}g_{\phi\phi}-g_{\psi\phi}\right]}}
\end{equation}
where $\gamma$ is the induced metric on the horizon, the integral is taken over the event horizon and it is assumed that the horizon is a hypersurface of constant $\rho$. This integral is problematic because of the terms in the square root, which make it difficult to compare with any corresponding expression derived by substituting in the charged metric coefficients. A more useful form for the induced metric is derived in Appendix A.

Re-writing the integral in terms of this new expression for the induced metric and transforming to the Einstein frame gives
\begin{equation}
A^E=\int{\sqrt{|\gamma^E|}}=\int{\frac{\sqrt{-g_{zz}h_1h_2}}{g_{tt}}\left(g_{t\psi}\sqrt{g_{t\phi}^2-g_{\phi\phi}g_{tt}}+g_{t\phi}\sqrt{g_{t\psi}^2-g_{\psi\psi}g_{tt}}\,\right)}
\end{equation}

Having re-expressed the area integrand in a more manageable form it is now possible to substitute for the charged metric coefficients obtained by comparing (\ref{genmet}) and (\ref{KKmetric}). After substituting for the various metric factors, the area of the charged metric becomes
\begin{equation}
\tilde A^E=\int{\sqrt{|\gamma^E|}\cosh{\alpha_1}\cosh{\alpha_2}}=A^E\cosh{\alpha_1}\cosh{\alpha_2}
\label{ADMA}
\end{equation}
It was shown in \cite{Horowitz:1993wt} that the horizon entropy is invariant under duality transformations and thus invariant for all varieties of string charge. This implies that the expression for the horizon area is definitive for all two charge metrics.

Having charged up the metric, it is necessary to calculate the conserved charges associated with the two 1-form fields $A^{(1)}$ and $A^{(2)}$. In general the gauge charges in five dimensions are given by
\begin{equation}
Q_i=\frac{1}{4\pi^2}\int_{S^3}{e^{-2\Phi_i}\star F^{(i)}}
\label{conchargeint}
\end{equation}
where $F=\ud A^{(i)}$ and the $e^{-2\Phi_i}$ factors, obtained by inspection of the Kaluza Klein reduced action given in (\ref{KKaction}), are
\begin{equation}
\Phi_1=\Phi+\frac{\sigma}{2} \hspace{4cm} \Phi_2=\Phi-\frac{3\sigma}{2}
\end{equation}

The integral given in (\ref{conchargeint}) has to be taken over a three sphere at infinity, so to simplify the algebra, it is convenient to work in spherical polar coordinates where $(\rho,z)\rightarrow(r,\theta)$ with $t$, $\psi$, and $\phi$ remaining unchanged. This means that the only pertinent component is $\star F^{(i)}_{\theta\psi\phi}$, since the integral has to be taken for a constant $t$ and $r$ cross-section. Furthermore, the components of $\star F^{(i)}$ only need to be known at asymptotic infinity, so only the first order terms of the Taylor expansion at infinity need to be considered. The $\star F^{(i)}_{\theta\psi\phi}$ component can immediately be reduced to the sum of three terms, by virtue of the metric having three Killing vectors and $g_{\theta t}=0$ since there is assumed to be no rotation in the $\theta$ direction. This gives
\begin{equation}
\star F^{(i)}_{\theta\psi\phi}=\frac{1}{\sqrt{|g|}}\left[g_{\theta\theta}g_{\psi\psi}g_{\phi\phi}\tilde\epsilon^{rt\theta\psi\phi}F_{rt}+ g_{\theta\theta}g_{\psi t}g_{\phi\phi}\tilde\epsilon^{r\psi\theta t\phi}F_{r\psi} + g_{\theta r}g_{\psi\psi}g_{\phi\phi}\tilde\epsilon^{\theta tr\psi\phi}F_{\theta t}\right]
\label{Fstar}
\end{equation}
where $\tilde\epsilon^{\mu\nu\rho\sigma\tau}$ is the Levi-Civita tensor density and $\tilde\epsilon^{tr\theta\psi\phi}=1$.

The leading order terms for the metric functions can be determined by considering the asymptotic expansion at infinity of the general spherical five dimensional metric given in \cite{Myers:1986un}. This series expansion indicates that the metric coefficients at infinity are unchanged by the process of adding charges to (\ref{genmet}), which allows (\ref{Fstar}) to be simplified further because it is now evident that the $g_{\psi t}$ and $g_{\theta r}$ coefficients are zero at infinity. Substituting the leading order terms for $g_{\psi\psi}$, $g_{\phi\phi}$, $g_{\theta\theta}$, and $g$ into (\ref{Fstar}) gives
\begin{equation}
\star F^{(i)}_{\theta\psi\phi}=-r^3\sin{\theta}\cos{\theta}F^{(i)}_{rt}
\label{Fstar2}
\end{equation}

To obtain an expression for $F_{rt}\equiv\partial_r A_t-\partial_t A_r$ at infinity, it is necessary to substitute for $g_{tt}$ from (\ref{gtt}) to give
\begin{equation}
A^{(i)}_t = \frac{8GM}{3\pi r^2+8GM\sinh^2{\alpha_i}}\sinh{\alpha_i}\cosh{\alpha_i}\stackrel{r=\infty}{\rightarrow}\frac{8GM}{3\pi r^2}\sinh{\alpha_i}\cosh{\alpha_i}
\end{equation}
This then allows $F_{rt}$ to be calculated
\begin{equation}
F^{(i)}_{rt}=-\frac{8GM}{3\pi r^3}\sinh{2\alpha_i}
\end{equation}
Putting this together with (\ref{Fstar2}) gives
\begin{equation}
\star F^{(i)}_{\theta\psi\phi}=\frac{4GM\sin{2\theta}}{3\pi}\sinh{2\alpha_i}
\end{equation}

The $e^{-2\Phi_i}$ factors will both go to one by virtue of them being functions of $h_i$, which go to one at infinity, as is easily verified by substituting for $g_{tt}$ from (\ref{gtt}) and taking the limit as $r\rightarrow\infty$. This now allows the integral given in (\ref{conchargeint}) to be evaluated, and the conserved charges to be calculated, as
\begin{equation}
Q_i=\frac{4GM}{3\pi}\sinh{2\alpha_i}
\label{ConsQ}
\end{equation}

\newsection{The Two Charge Dual Rotating Ring}
\label{sec:TwoChargeSol}
\newsubsection{Neutral Dual Rotating Ring}
\label{sec:neutralring}

The dual spinning black ring was first presented in a paper by Pomeransky and Sen'kov \cite{Pomeransky:2006bd} for the equilibrium ring with a mostly negative signature. The form of the metric used in this paper differs from their version as it has a mostly positive signature, as well as a few other differences. The coordinates $\phi$ and $\psi$ have been interchanged so that $\psi$ corresponds to angles on the $S^1$ in accordance with the single spinning ring. The constant $k$ has also been rescaled so that when $\nu\rightarrow 0$, $k$ is equivalent to the ring radius parameter $R$ used in \cite{Emparan:2006mm}. A side effect of the re-scaling of $k$ is that the values given below for the thermodynamic properties will disagree by some power of $\sqrt{2}$ with equivalent values presented elsewhere\footnote{To convert between the values presented here and those in \cite{Elvang:2007hs} it is necessary to replace each instance of $k$ in this paper with $k\sqrt{2}$}. Bearing all this in mind, the dual rotating ring metric is now given by
\begin{eqnarray}
ds^2 & = &-\frac{H(y,x)(\ud t+\Omega)^2}{H(x,y)}-\frac{F(x,y)\ud\psi^2}{H(y,x)}-\frac{2J(x,y)\ud\phi\ud\psi}{H(y,x)} \nonumber \\
&& +\frac{F(y,x)\ud\phi^2}{H(y,x)} +\frac{k^2H(x,y)}{(x-y)^2(1-\nu)^2}\left(\frac{\ud x^2}{G(x)}-\frac{\ud y^2}{G(y)}\right)
\label{neutralring}
\end{eqnarray}
where
\begin{eqnarray}
\Omega &= &-\frac{k\lambda\sqrt{2(1+\nu)^2-2\lambda^2}}{H(y,x)}\Biggl[(1-x^2)y\sqrt{\nu}\ud\phi \nonumber \\
&& \hspace{4.2cm} +\frac{(1+y)\left[1+\lambda-\nu+x^2y\nu(1-\lambda-\nu)+2\nu x(1-y)\right]\ud \psi}{(1-\lambda+\nu)}\Biggr] \\
G(x) &= &(1-x^2)(1+\lambda x+\nu x^2) \\
H(x,y) &= &1+\lambda^2-\nu^2+2\lambda\nu(1-x^2)y+2x\lambda(1-y^2\nu^2)+x^2y^2\nu(1-\lambda^2-\nu^2) \\
J(x,y) &= & \frac{k^2(1-x^2)(1-y^2)\lambda\sqrt{\nu}\left[1+\lambda^2-\nu^2+2(x+y)\lambda\nu-xy\nu(1-\lambda^2-\nu^2)\right]}{(x-y)(1-\nu)^2} \\
F(x,y) &= & \frac{k^2}{(x-y)^2(1-\nu)^2}\Biggl[G(x)(1-y^2)\left\{[(1-\nu)^2-\lambda^2](1+\nu)+y\lambda[1-\lambda^2+2\nu -3\nu^2]\right\} \nonumber \\
&& \hspace{3.1cm} \, + G(y)\left\{ 2\lambda^2+x\lambda[(1-\nu)^2+\lambda^2]+x^2[(1-\nu)^2-\lambda^2](1+\nu) \right. \nonumber \\
&& \hspace{3.1cm} \, \left. +x^3\lambda(1-\lambda^2-3\nu^2+2\nu^3)-x^4(1-\nu)\nu(\lambda^2+\nu^2-1)\right\}\Biggr]
\end{eqnarray}
This form of the metric has been specially chosen so that when $\nu\rightarrow 0$ the singly rotating equilibrium ring given by Emparan and Reall in \cite{Emparan:2006mm} is recovered with $\lambda$ replacing $\nu$.

Apart from the parameter re-scaling, the other properties of the dual rotating ring metric are identical to those given in the literature i.e. the horizons are at
\begin{equation}
y_h=\frac{-\lambda\pm\sqrt{\lambda^2-4\nu}}{2\nu}
\label{ringhorizon}
\end{equation}
where $y$ is restricted such that $-\infty<y<-1$ and the ring parameters $\lambda$ and $\nu$ are constrained as $0\le\nu<1$ and $2\sqrt{\nu}\le\lambda<1+\nu$.

As for the singly spinning ring, infinity is where $x=y=-1$ and the centre of the ring is at $x=1$, $y=-1$. A more detailed indication of the contour lines of constant $x$ and $y$ is given in Appendix B. The dual rotating ring is asymptotically flat in the limit where $x=y=-1$ permitting the asymptotic coordinates \cite{Elvang:2007hs}
\begin{equation}
x=-1+\frac{4k^2}{r^2}\alpha^2\cos^2{\theta} \hspace{3cm}
y=-1-\frac{4k^2}{r^2}\alpha^2\sin^2{\theta}
\label{asympcoords}
\end{equation}
where
\begin{equation}
\alpha=\sqrt{\frac{1+\nu-\lambda}{1-\lambda}}
\end{equation}
In these coordinates $0\le r<\infty$ and $0\le\theta\le 2\pi$ and asymptotic infinity is approached as $r\rightarrow\infty$.

Applying the coordinate transformations given by (\ref{asympcoords}) to (\ref{neutralring}) gives the physical properties of the dual rotating ring
\begin{eqnarray}
M &= &\frac{3\pi\lambda k^2}{2G(1-\lambda+\nu)} \\
J_\psi &= &\frac{\pi\lambda k^3(1+\lambda-6\nu+\nu^2+\lambda\nu)\sqrt{2}\sqrt{(1+\nu)^2-\lambda^2}}{2G(1-\nu)^2(1-\lambda+\nu)^2} \\
J_\phi &= &\frac{\pi\lambda k^3\sqrt{2\nu}\sqrt{(1+\nu)^2-\lambda^2}}{G(1-\nu)^2(1-\lambda+\nu)}
\end{eqnarray}
and the area of the event horizon is given by
\begin{equation}
A = \frac{8\sqrt{2}\pi^2k^3\lambda y_h(1+\lambda+\nu)}{(1-\nu)^2(1-y_h^2)}
\end{equation}
\newsubsection{[F(w),P] Dual Rotating Ring Metric}

The full 10D type IIB string solution for the dual rotating metric, after substituting in (\ref{chgmet}), is given by
\begin{eqnarray}
ds_6^2 &= & -\frac{m_2(y,x)}{m_1(x,y)}\ud t^2 - \frac{H(y,x)\left( \cosh{\alpha_2}\ud t + \sinh{\alpha_2}\ud w\right)^2}{m_1(x,y)} \nonumber \\
&& + \frac{H(y,x)\left( \cosh{\alpha_2}\ud t + \sinh{\alpha_2}\ud w +\Omega\cosh{\alpha_1}\right)^2}{m_1(x,y)} \nonumber \\
&& -\frac{F(x,y)}{H(y,x)}\ud \psi^2 - 2\frac{J(x,y)}{H(y,x)}\ud \phi \ud \psi +\frac{F(y,x)}{H(y,x)}\ud \phi^2 + \frac{k^2H(x,y)}{(x-y)^2(1-\nu)^2}\left(\frac{\ud x^2}{G(x)}-\frac{\ud y^2}{G(y)}\right) \nonumber \\
&& + \frac{m_2(x,y)}{m_1(x,y)}\ud w^2 + \frac{2\cosh{\alpha_2}\sinh{\alpha_2}\left[H(y,x)-H(x,y)\right]}{m_1(x,y)}\ud t\ud w
\end{eqnarray}
where
\begin{equation}
m_n(x,y) = H(y,x)\sinh^2{\alpha_n}-H(x,y)\cosh^2{\alpha_n}
\end{equation}
The four additional dimensions have been suppressed for brevity, but they are given by $g_{\mu\nu}=\delta_{\mu\nu}$ where $\mu,\nu=6,7,8,9$. The additional fields, introduced through the duality process are given by
\begin{eqnarray}
B^{(6)}_{t\phi} &= & \frac{\sqrt{2}k\lambda y\sqrt{\nu}\sinh{\alpha_1}\sinh{\alpha_2}\sqrt{1+\lambda+\nu}\sqrt{1-\lambda+\nu}(x^2-1)}{m_1(x,y)} \\
B^{(6)}_{t\psi} &= & \frac{\sqrt{2}k\lambda\sqrt{1+\lambda+\nu}(1+y)\left[(\lambda+\nu-1)(yx^2\nu-1)+2\nu(1-x+xy)-2\right]\sinh{\alpha_1}\sinh{\alpha_2}}{m_1(x,y)\sqrt{1-\lambda+\nu}} \\
B^{(6)}_{tw} & =& -\frac{\sinh{\alpha_1}\cosh{\alpha_1}\left[H(y,x)-H(x,y)\right]}{m_1(x,y)} \\
B^{(6)}_{\phi w} &= & -\frac{\sqrt{2}k\lambda y\sqrt{\nu}\sinh{\alpha_1}\cosh{\alpha_2}\sqrt{1+\lambda+\nu}\sqrt{1-\lambda+\nu}(x^2-1)}{m_1(x,y)} \\
B^{(6)}_{\psi w} &= &  - \frac{\sqrt{2}k\lambda(1+y)\sqrt{1+\lambda+\nu}\left[(\lambda+\nu-1)(yx^2\nu-1)+2\nu(1-x+xy)-2\right]\cosh{\alpha_2}\sinh{\alpha_1}}{m_1(x,y)\sqrt{1-\lambda+\nu}}
\end{eqnarray}
and
\begin{equation}
e^{-2\Phi}=-\frac{m_1(x,y)}{H(x,y)}
\end{equation}
In this metric, the canonical coordinates have been replaced with the toroidal ($x$, $y$) coordinates\footnote{The transformations used can be found in \cite{Pomeransky:2006bd}} which may be concerning, since the derivation of the previous section was in terms of the ($\rho$, $z$) coordinates. Fortunately, all of the transformations used to charge up the generic metric were independent of these coordinates, so they can be transformed with impunity.

In order to calculate the various physical properties of the charged ring it is necessary to now reduce the metric back down to five dimensions. This has already been done in (\ref{KKmetric}), so substituting for the various metric coefficients and transforming to the Einstein frame gives\footnote{Inspection of the $\Omega$ function shows that this metric is free from Dirac-Misner singularities.}
\begin{eqnarray}
ds_5^2 &= & -\frac{\left[\left(H(y,x)-H(x,y)\right)\cosh{\alpha_2}\sinh{\alpha_2}\ud t+H(y,x)\sinh{\alpha_2}\cosh{\alpha_1}\Omega\right]^2}{(m_1(x,y)m_2(x,y))^{2/3}H(x,y)^{1/3}} \nonumber \\
&& -\left(\frac{m_2(x,y)}{m_1(x,y)^2H(x,y)}\right)^{1/3}\left(m_2(y,x)\ud t^2 + H(y,x)\Omega^2\cosh^2{\alpha_1} + 2H(y,x)\cosh{\alpha_1}\cosh{\alpha_2}\Omega\ud t\right) \nonumber \\
&& - \left(\frac{m_1(x,y)m_2(x,y)}{H(x,y)}\right)^{1/3}\left[\frac{F(x,y)}{H(y,x)}\ud \psi^2 - 2\frac{J(x,y)}{H(y,x)}\ud \phi \ud \psi +\frac{F(y,x)}{H(y,x)}\ud \phi^2 \right. \nonumber \\
&& \left. + \frac{k^2H(x,y)}{(x-y)^2(1-\nu)^2}\left(\frac{\ud x^2}{G(x)}-\frac{\ud y^2}{G(y)}\right)\right]
\label{FPring}
\end{eqnarray}
The 1-form gauge fields are now given by
\begin{eqnarray}
A_t^{(1)} &= &\frac{\left[H(y,x)-H(x,y)\right]\cosh{\alpha_2}\sinh{\alpha_2}}{m_2(x,y)} \\
A_\psi^{(1)} &= &-\frac{k\lambda\sqrt{2(1+\nu)^2-2\lambda^2}(1+y)\left[1+\lambda-\nu+x^2y\nu(1-\lambda-\nu)+2\nu x(1-y)\right]\sinh{\alpha_2}\cosh{\alpha_1}}{(1-\lambda+\nu)m_2(x,y)} \\
A_\phi^{(1)} &= &-\frac{k\lambda\sqrt{2(1+\nu)^2-2\lambda^2}(1-x^2)y\sqrt{\nu}\sinh{\alpha_2}\cosh{\alpha_1}}{m_2(x,y)} \\
A_t^{(2)} &= &-\frac{\left[H(y,x)-H(x,y)\right]\cosh{\alpha_1}\sinh{\alpha_1}}{m_1(x,y)} \\
A_\psi^{(2)} &= & - \frac{\sqrt{2}k\lambda(1+y)\sqrt{1+\lambda+\nu}\left[(\lambda+\nu-1)(yx^2\nu-1)+2\nu(1-x+xy)-2\right]\cosh{\alpha_2}\sinh{\alpha_1}}{m_1(x,y)\sqrt{1-\lambda+\nu}} \\
A_\phi^{(2)} &= &-\frac{\sqrt{2}k\lambda y\sqrt{\nu}\sinh{\alpha_1}\cosh{\alpha_2}\sqrt{1+\lambda+\nu}\sqrt{1-\lambda+\nu}(x^2-1)}{m_1(x,y)}
\end{eqnarray}
The 2-form field is given by
\begin{eqnarray}
B_{t\phi} &= & \frac{\sqrt{2}k\lambda y\sqrt{\nu}\sinh{\alpha_1}\sinh{\alpha_2}\sqrt{1+\lambda+\nu}\sqrt{1-\lambda+\nu}(x^2-1)}{m_1(x,y)} \\
B_{t\psi} &= & \frac{\sqrt{2}k\lambda\sqrt{1+\lambda+\nu}(1+y)\left[(\lambda+\nu-1)(yx^2\nu-1)+2\nu(1-x+xy)-2\right]\sinh{\alpha_1}\sinh{\alpha_2}}{m_1(x,y)\sqrt{1-\lambda+\nu}}
\end{eqnarray}
and the scalar functions are given by
\begin{equation}
e^{-2\Phi}=-\frac{m_1(x,y)}{H(x,y)} \hspace{3cm} e^{2\sigma}=\frac{m_2(x,y)}{m_1(x,y)}
\end{equation}

\newsection{Physical Properties of the Generic Charged Metric}
\label{PhysicalProperties}

Having obtained the metric for the [F(w),P] charged black ring (\ref{FPring}) and the more general two charge metric (\ref{KKmetric}), it is now possible to work out some of the physical properties of these solutions. In fact, most of the distinguishing features of the charged solutions are the same as for the neutral solution: the $x$, $y$ (or $\rho$, $z$ for the general solution), $\psi$, and $\phi$ coordinates vary over the same ranges, any physical constraints on the original neutral metric will be unchanged e.g. the limits on $\lambda$ and $\nu$ are exactly the same for the two charge black ring, and the horizons will still be given by $g^{\rho\rho}=0$ or in the case of the charged ring (\ref{ringhorizon}). The reason that these properties are unchanged for the charged solution is because they all depend, to some extent, upon the $g_{\rho\rho}$ and $g_{zz}$ coefficients of the metric, which are unaffected by the boost and T-duality transformations. The addition of the charges does have some effect on the ADM mass, angular momenta, and the area but these have all been calculated in (\ref{ADMM}), (\ref{ADMJ}), and (\ref{ADMA}) respectively. The conserved gauge charges for the charged metric have also been calculated in (\ref{ConsQ}) and are explicitly given by
\begin{equation}
Q_1=\frac{4G}{3\pi}M_0\sinh{2\alpha_1} \hspace{3cm} Q_2=\frac{4G}{3\pi}M_0\sinh{2\alpha_2}
\end{equation}
It is worth noting that the charges are directly related to their respective boosts, which verifies the physical picture of the linear momentum being exchanged for winding charges when the metric is T-dualised.

\begin{figure}[htbp]
\center
\includegraphics[viewport=164 279 635 770,width=8cm,clip]{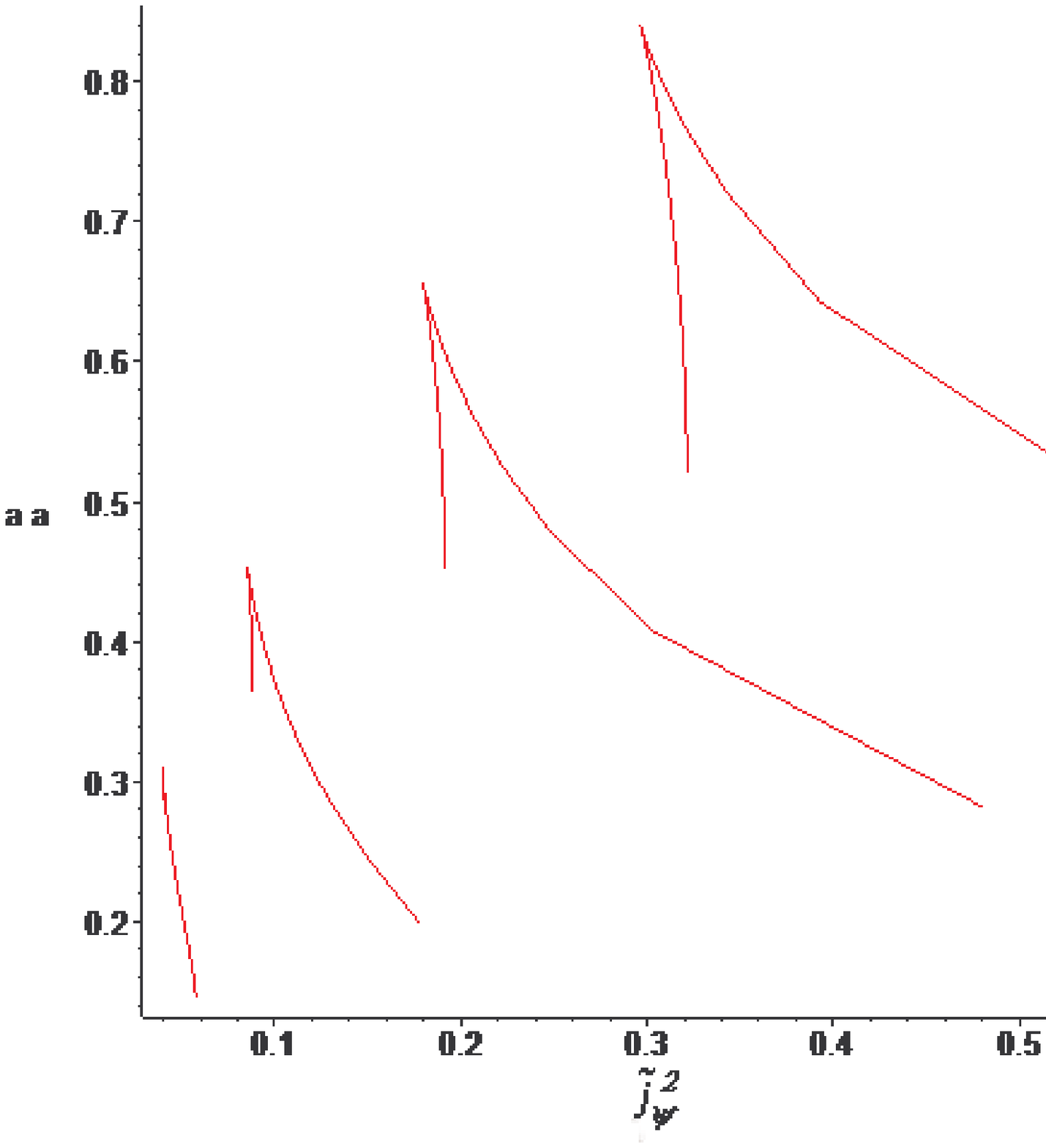}
\caption{This gives an example of how the $a$ vs ${j_\psi}^2$ plots change for the dual spinning black ring as the charge is increased, with $\alpha_1=\alpha_2=1,\frac{3}{4},\frac{1}{2},\frac{3}{10}$ from the top right to the bottom left respectively.}
\label{fig:avsjpsi}
\end{figure}

To get an idea of how the behaviour of the charged solution differs from that of the neutral ring, it is a good idea to plot some phase diagrams showing how the physical properties, like the angular momentum and horizon area, vary with the charge. To this end, the charged dual rotating ring solution given in (\ref{FPring}) will be considered. The phase space of the dual rotating black ring has been fairly extensively studied in \cite{Elvang:2007hs} and \cite{Emparan:2008eg}, so the following discussion will concentrate on where the behaviour of the charged dual rotating ring (and hence the more general solution given by (\ref{KKmetric})) departs from that of the neutral ring.

Before plotting the various physical properties, it is beneficial to re-define them so they are scale independent. The obvious candidate for fixing the scale is the ADM mass, so expressing the angular momentum and horizon area in terms of this, along with the conventional normalisation, gives the following relations
\begin{equation}
j^2=\frac{27\pi J^2}{32GM^3} \hspace{3cm} a_H=\frac{3}{16}\sqrt{\frac{3}{\pi}}\frac{A}{(GM)^{3/2}}
\label{dimtrans}
\end{equation}
The square of the angular momentum is given above because it is more convenient to plot in terms of $j^2$, since it is always positive. For the neutral ring, $j_\psi$ and $j_\phi$ are constrained such that
\begin{equation}
j_\phi\le\frac{1}{4} \hspace{3cm} j_\psi\ge\frac{3}{4}
\end{equation}
This means that the angular momenta can never be equal and $j_\phi/j_\psi\le 1/3$ for all permissible values of $\nu$ and $\lambda$. The constraints on $j_\psi$ and $j_\phi$ are dependent on the form of the metric coefficients, with the constraints on the angular momenta for the dual rotating ring being a consequence of the restrictions on $\lambda$ and $\nu$. In general these restrictions will always have to be calculated for each given metric.

As can be seen by examining (\ref{ADMJ}) and (\ref{ADMA}), the only difference between the angular momenta and area of the neutral metric and the charged metric is a factor of $\cosh{\alpha_1}\cosh{\alpha_2}$, which is the same for $J_\psi$, $J_\phi$, and $A$. When these are combined with the ADM mass to give the dimensionless quantities of (\ref{dimtrans}), the relationship between the neutral metric physical properties and the charged metric physical properties is
\begin{equation}
\{\tilde j,\tilde a\} =\frac{2\sqrt{2}\cosh{\alpha_1}\cosh{\alpha_2}}{\left(\cosh{2\alpha_1}+\cosh{2\alpha_2}\right)^{(3/2)}}\{j,a\}
\label{chargeneutral}
\end{equation}
where the tilde denotes the charged angular momentum and area. Unfortunately, because the factors are all equal for the various quantities they all tend to cancel out, meaning that the physics of the charged metric is very similar to that of the neutral metric. This is exemplified by the fact that the maximum and minimum of $\tilde j$ is exactly the same as for the neutral metric, no matter how large the charges are. This is because the multiplying factor varies between $0$ and $1$.

The multiplying factor in (\ref{chargeneutral}) encodes all of the differences between the properties of the charged metric and the neutral metric. The denominator of the multiplying factor in (\ref{chargeneutral}) is larger than the terms in the numerator for all $|\alpha_i|>0$ so this factor has a maximum of 1 for $\alpha_1=\alpha_2=0$ and then exponentially decays toward 0 as the charges are increased. This means that the charged metric angular momenta and horizon area will always be smaller than the corresponding neutral metric if only the charges are varied. In the case of the dual rotating black ring, this agrees with the intuitive interpretation of the ring being balanced by the charge as well as the angular momentum. The charge helps to balance the tension trying to collapse the ring and thus allows a ring that would otherwise be unstable, if only balanced by the centrifugal force, to remain in equilibrium.

The form of the expressions for the angular momentum shows that for any given ring, as the charge is increased the angular momentum in the $\phi$ and the $\psi$ plane will have to decrease for the ring to remain in equilibrium, with the speed of the rotation decreasing as the charge increases. Unfortunately, since the multiplying factor in (\ref{chargeneutral}) only asymptotically approaches zero, there is no way that the ring can only be balanced by rotation in the $\phi$ direction, or by the charge alone. In order for the angular momentum in either direction to reach zero, the charge would have to be infinite.

Figure \ref{fig:avsjpsi} shows how the ring area $a$ varies with $\tilde j_\psi^2$ for various different values of $\alpha_1$ and $\alpha_2$. The curves are constrained so that ${\tilde j_\phi}^2={1/500}$.
The effect of increasing or decreasing $\alpha_1$ or $\alpha_2$ is to move the phase curve closer or further away from the origin respectively. It doesn't matter whether $\alpha_1$ or $\alpha_2$ is varied, since the factor in (\ref{chargeneutral}) is symmetrical under interchange of $\alpha_1$ and $\alpha_2$. The basic effect of varying the charge is to replicate the phase curve of a neutral ring with larger $j_\psi^2$ other than that, the phase curves are identical in shape to those of the neutral ring. This behaviour can be generalised to any other metric given by (\ref{KKmetric}) but, obviously, the physical interpretation would depend upon the physics of the neutral metric (\ref{genmet}).

To find the point where the black ring angular momentum in the $\psi$ direction is minimized and the area maximized it is necessary to deploy a Lagrange multiplier to fix $\tilde j_\phi$ whilst $\tilde j_\psi$ is minimized. Doing this gives the value of $\lambda$ in terms of $\nu$ where $\tilde j_\psi$ and hence $a$ are maximized. Unsurprisingly, this gives exactly the same expression as for the neutral ring, where
\begin{equation}
\lambda=\frac{1}{4}\left(-1-\nu+\sqrt{(9+\nu)(1+9\nu)}\right)
\end{equation}
This is because the form of $\tilde j_\psi$ and $\tilde j_\phi$ is exactly the same, so any factors that are introduced by $\alpha_i\ne 0$ cancel out completely.

\begin{figure}[htbp]
\center
\includegraphics[viewport=235 375 563 700,width=8cm,clip]{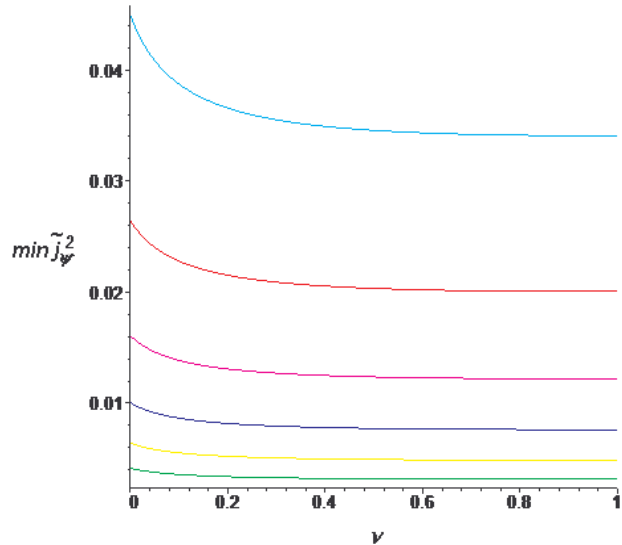}
\caption{This plot shows how the minimum angular momentum along the $S^1$ reduces as the charge is increased from $\alpha_1=\alpha_2=1$ through to $\alpha_1=\alpha_2=2$ in increments of 0.2. The highest curve represents the smallest total charge and the lowest curve represents the largest total charge.}
\label{fig:minjpsivnu}
\end{figure}

Figure \ref{fig:minjpsivnu} gives some examples of how the ring angular momentum ${\tilde j_\psi}^2$ decreases as $\nu\rightarrow 1$ i.e. as the rings get fatter, and decreases for all $\nu$ as the charge is increased. Although the minimum angular momentum curves for larger values of $\alpha_1$ and $\alpha_2$ seem to be approaching $\tilde j_\psi=0$ rapidly, they will only ever asymptotically approach it. This means that even though the addition of a small charge will significantly decrease the angular momentum needed for the ring to remain balanced, it will never sustain a balanced ring without angular momentum in at least the $\psi$ direction. The variation of the angular momentum with the charge is similar for a generic metric but the form of the plots will obviously vary.

\begin{figure}[htbp]
\center
\includegraphics[viewport=228 360 564 680,width=8cm,clip]{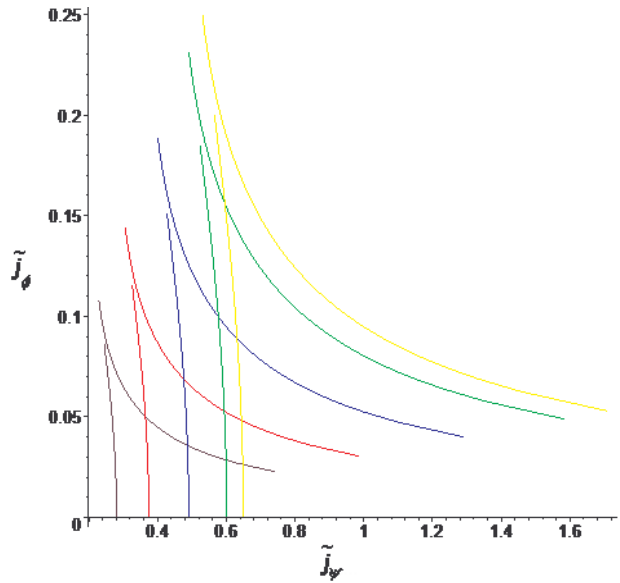}
\caption{This plot gives the lines delimiting the phase space of the charged ring when the charges are equal for $\alpha_1=\alpha_2=0,0.2,0.4,0.6,0.8$ from top to bottom respectively.}
\label{fig:jpsivjphi}
\end{figure}

Figure \ref{fig:jpsivjphi} shows the phase space of the charged black ring for $\tilde j_\phi$ against $\tilde j_\psi$ for various values of $\alpha_1$ and $\alpha_2$ ranging from 0 to 0.8. For any given charge, the permissible values of $\tilde j_\psi$ and $\tilde j_\phi$ are those that are enclosed by the curve and the $\tilde j_\psi$ axis. Values of $\tilde j_\psi$ and $\tilde j_\phi$ outside of this enclosed region are not physically permissible. As with all the other plots, the increased charge has the effect of decreasing the angular momentum, so the larger the charge the smaller the range of the permissible angular momenta in the $\psi$ and $\phi$ directions.

To analyse the extremal doubly spinning ring, it is necessary to substitute $\lambda=2\sqrt{\nu}$. For this case, the analysis of the charged version proceeds in a similar manner to that of the non-extremal doubly spinning ring. The mass, angular momentum and horizon area are still given by the expressions in (\ref{ADMM}), (\ref{ADMJ}), and (\ref{ADMA}) which differ from the uncharged versions by functions only involving $\cosh{\alpha_1}$ and $\cosh{\alpha_2}$. Thus all of the physical properties, that depend upon the charge, will vary in the same way as previously described.

\newsection{Conclusions}

This paper was principally concerned with constructing a two charge supersymmetric string solution given an initial solution to Einstein's field equations. It then used the dual rotating black ring found by Pomeransky and Sen'kov in \cite{Pomeransky:2006bd} as an example of how the general method can be applied to obtain specific solutions. The generic charged metric and its physical properties were derived in section \ref{Chargemetsec}, before going on to present the dual spinning black ring metric at the beginning of section \ref{sec:neutralring}. This allowed the various physical properties of the charged metric to be examined using the dual rotating black ring as an example.

Section \ref{Theory} explained the theory behind converting the original five dimensional solution to Einstein's equations into a form that is suitable to be manipulated using the string theory techniques, as well as setting out the required formalism. The technique involves lifting the metric to ten dimensions, boosting the metric, T-dualising to obtain a fundamental charge, and then boosting again to produce a second charge. The resulting solution is then Kaluza-Klein compactified back down to five dimensions, so that the physical properties of the solution can be explored. The first part of section \ref{Chargemetsec} defined a generic starting metric and then applied the above techniques to generate a new solution with fundamental and momentum string charges.

The second part of section \ref{Chargemetsec} derived the physical properties of the charged metric in terms of those of the neutral metric. The analysis showed that the ADM mass, angular momentum, and the horizon area of the charged metric were directly proportional to the respective quantities for the neutral metric with the multiplying factor being a function of the boost parameters used to generate the charged solution. The derivation of the conserved charges showed that the gauge charges were related to the boost parameters by a factor of $\sinh{2\alpha}$, where $\alpha$ is the boost parameter.

Section \ref{sec:TwoChargeSol} presented the neutral dual rotating black ring solution, along with some of the associated formalism. This consisted of the thermodynamic quantities of the black hole, along with a brief description of the parameters and the toroidal coordinates used to describe the solution. The second part of this section applied the charging technique to the dual rotating black ring to obtain the five dimensional charged metric in the Einstein frame, along with all its associated fields. This involved deriving the ADM mass, angular momenta, event horizon area, and canonical charges.

Section \ref{PhysicalProperties} looked at how the addition of the string charges varied the physical properties of the generic black hole solution, and the dual rotating black ring in particular. It was found that the addition of the charges had little effect overall with the angular momenta and area being the only properties that were affected. In general the area and angular momenta were reduced as the charges were increased but the angular momentum could never be decreased to zero for finite charge. This is because the area and angular momenta of the charged solution only differed from those of the neutral solution by a multiplying factor which was a function of the boost parameters. This multiplying factor decayed exponentially from $1$ when the charges were set to zero and asymptotically approached $0$ for large values of the boost parameters.

It was shown that the behaviour of the charged black ring was virtually identical to that of the neutral ring, with the charge playing a similar role to that of the angular momentum in the neutral case. This was as expected but it was shown that the charged ring must have non-zero angular momentum in the $\psi$ and $\phi$ direction to remain balanced. As the charge is increased, the angular momentum required to keep the ring in equilibrium decreases exponentially but only ever asymptotically approaches zero. This was born out by the phase plots of the charged black ring angular momenta, which were identical in form to those of the neutral black ring but were shrunk by a factor depending upon the charges.

This work looked at the charged metric when fundamental string and momentum charges were added but it is possible to generate a $[D1,D5]$ charged solution by carrying out some further duality transformations. It would be interesting to see whether it would be possible to extend the methods used in this paper to charge up the generic metric, given at the beginning of section \ref{Chargemetsec}, to produce a generic $[D1,D5]$ solution. It is possible to do this when the specific form of the metric coefficients are known (such as in \cite{Elvang:2003mj}) but this requires some of the 2-forms, generated from the original metric coefficients, to be differentiated and integrated. Given this, it is not immediately obvious whether this method could be applied to a generic metric solution. The $[D1,D5]$ solution would have exactly the same physical properties \cite{Horowitz:1993wt} as the solution presented here, but it would provide a broader basis for the investigation of the microscopic entropy of black hole solutions.

\newsection{Appendix A: Derivation of the Induced Metric on the Event Horizon for a Generic Metric}

The relationship between the area of the charged metric and that of the neutral metric appears to be simple but substitution of the charged metric coefficients in terms of the neutral coefficients doesn't immediately give the desired relationship between the two horizon areas. Instead, consider the determinant of the neutral metric (\ref{genmet})
\begin{equation}
\left|
\begin{array}{ccccc}
   g_{tt}    & g_{t\psi}    & g_{t\phi}    & 0            & 0      \\
   g_{t\psi} & g_{\psi\psi} & g_{\psi\phi} & 0            & 0      \\
   g_{t\phi} & g_{\psi\phi} & g_{\phi\phi} & 0            & 0      \\
   0         &      0       &      0       & g_{\rho\rho} & 0      \\
   0         &      0       &      0       & 0            & g_{zz}
\end{array}
\right|=\Lambda(\rho,z)
\end{equation}
where $\Lambda(\rho,z)$ is an arbitrary function depending on the metric coefficients. It is only a function of the variables $\rho$ and $z$, since the other coordinates can't appear in the metric coefficients by virtue of their being Killing vectors.

A little manipulation shows that the $g_{\rho\rho}$ and $g_{zz}$ coefficients can be factored out to give
\begin{equation}
g_{\rho\rho}g_{zz}\left|
\begin{array}{ccc}
   g_{tt}    & g_{t\psi}    & g_{t\phi} \\
   g_{t\psi} & g_{\psi\psi} & g_{\psi\phi} \\
   g_{t\phi} & g_{\psi\phi} & g_{\phi\phi}
\end{array}
\right|=\Lambda(\rho,z)
\end{equation}
This can then be rearranged to give
\begin{equation}
\left|
\begin{array}{ccc}
   g_{tt}    & g_{t\psi}    & g_{t\phi} \\
   g_{t\psi} & g_{\psi\psi} & g_{\psi\phi} \\
   g_{t\phi} & g_{\psi\phi} & g_{\phi\phi}
\end{array}
\right|=\frac{\Lambda(\rho,z)}{g_{\rho\rho}g_{zz}}=\frac{g^{\rho\rho}\Lambda(\rho,z)}{g_{zz}}
\end{equation}
where it is assumed that the event horizon is a hypersurface of constant $\rho$. In this case $g^{\rho\rho}=0$ so, after expanding the determinant, the above equation gives an identity relating some of the coefficients in the neutral metric.
\begin{equation}
g_{tt}(g_{\psi\psi}g_{\phi\phi}-g_{\psi\phi}^2)-g_{t\psi}^2g_{\phi\phi}+2g_{t\psi}g_{t\phi}g_{\psi\phi}-g_{t\phi}^2g_{\psi\psi}=0
\label{metid}
\end{equation}

The main difficulty in calculating the area in terms of the metric functions is due to the square root in the integral. The only non-trivial way to eliminate this is to express the determinant of the induced metric in a manifestly squared form. To do this complete the square on the metric given in (\ref{genmet})
\begin{equation}
ds^2=g_{tt}\left(dt+\frac{g_{t\psi}\ud\psi+g_{t\phi}\ud\phi}{g_{tt}}\right)^2-\frac{(g_{t\psi}\ud\psi+g_{t\phi}\ud\phi)^2}{g_{tt}}+2g_{\psi\phi}\ud\psi\ud\phi + g_{\psi\psi}\ud\psi^2+g_{\phi\phi}\ud\phi^2+g_{\rho\rho}\ud \rho^2 + g_{zz}\ud z^2
\end{equation}

Armed with this expression and the identity given in (\ref{metid}), it is now possible to express the determinant of the induced metric in a manifestly squared form
\begin{eqnarray}
\gamma & = & g_{zz}\left(g_{\psi\psi}g_{\phi\phi}-g_{\psi\phi}^2\right) \nonumber \\
			 & \equiv & g_{zz}\left[\frac{g_{t\psi}^2g_{t\phi}^2}{g_{tt}^2}-\frac{g_{t\psi}^2}{g_{tt}}\left(\frac{g_{t\phi}^2}{g_{tt}}-g_{\phi\phi}\right)-\frac{g_{t\phi}^2}{g_{tt}}\left(\frac{g_{t\psi}^2}{g_{tt}}-g_{\psi\psi}\right)+\left(\frac{g_{t\phi}^2}{g_{tt}}-g_{\phi\phi}\right)\left(\frac{g_{t\psi}^2}{g_{tt}}-g_{\psi\psi}\right) \right. \nonumber \\
&& \left. \hspace{0.8cm} -\frac{g_{t\psi}^2g_{t\phi}^2}{g_{tt}^2}+2\frac{g_{t\psi}g_{t\phi}}{g_{tt}}\left(\frac{g_{t\psi}g_{t\phi}}{g_{tt}}-g_{\psi\phi}\right)-\left(\frac{g_{t\psi}g_{t\phi}}{g_{tt}}-g_{\psi\phi}\right)^2\right]
\end{eqnarray}
Now consider
\begin{equation}
\left(\frac{g_{t\phi}^2}{g_{tt}}-g_{\phi\phi}\right)\left(\frac{g_{t\psi}^2}{g_{tt}}-g_{\psi\psi}\right)-\left(\frac{g_{t\psi}g_{t\phi}}{g_{tt}}-g_{\psi\phi}\right)^2=(g_{\psi\psi}g_{\phi\phi}-g_{\psi\phi}^2)-\frac{g_{t\psi}^2g_{\phi\phi}-2g_{t\psi}g_{t\phi}g_{\psi\phi}+g_{t\phi}^2g_{\psi\psi}}{g_{tt}}
\end{equation}
which goes to zero by virtue of (\ref{metid}). It is also implicitly assumed that the particular metric under consideration represents a rotating black hole, so that $g^{tt}\not\rightarrow 0$ and hence $g_{tt}\not\rightarrow\infty$.

Simplifying and expanding the remaining terms for $\gamma$ gives
\begin{equation}
\gamma=-\frac{g_{zz}}{g_{tt}^2}\left(g_{t\psi}\sqrt{g_{t\phi}^2-g_{\phi\phi}g_{tt}}+g_{t\phi}\sqrt{g_{t\psi}^2-g_{\psi\psi}g_{tt}}\,\right)^2
\end{equation}
where (\ref{metid}) has been used again to give
\begin{equation}
-g_{\psi\psi}g_{t\phi}^2g_{tt}-g_{\phi\phi}g_{t\psi}^2g_{tt}+g_{\phi\phi}g_{\psi\psi}g_{tt}^2=g_{\psi\phi}^2g_{tt}^2-2g_{\psi\phi}g_{t\psi}g_{t\phi}g_{tt}
\end{equation}

This then implies
\begin{equation}
\sqrt{|\gamma|}=\frac{\sqrt{-g_{zz}}}{g_{tt}}\left(g_{t\psi}\sqrt{g_{t\phi}^2-g_{\phi\phi}g_{tt}}+g_{t\phi}\sqrt{g_{t\psi}^2-g_{\psi\psi}g_{tt}}\,\right)
\end{equation}
which is a much more useful form for comparing the neutral metric area with the charged metric area.

\begin{figure}[htbp]
\center
\includegraphics[viewport=5 40 245 480,width=8cm,angle=270,clip]{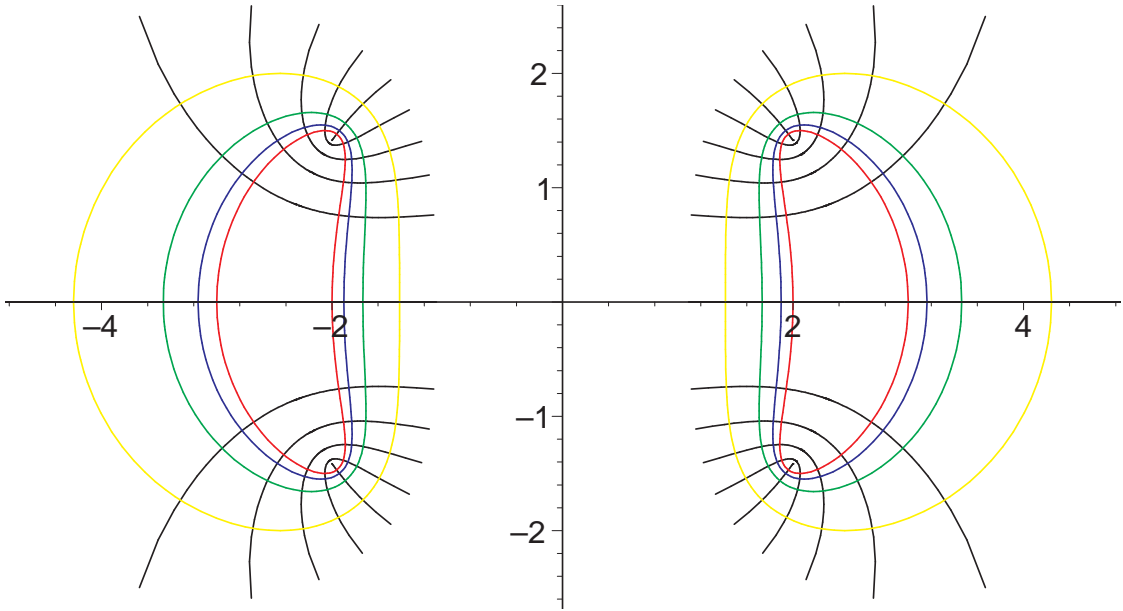}
\caption{A two dimensional cross-section of constant $\phi$ and $\psi$ (as well as the antipodal points $\phi+\pi$ and $\psi+\pi$) of the $(x,y)$ coordinates. The red through yellow lines show the lines of constant $y$ from -5 through to -2 respectively and the black lines show the $x$ contours. The vertical axis on this plot corresponds to asymptotic infinity at $y=-1$ and the horizontal axis corresponds to $x=\pm 1$.}
\label{fig:contours}
\end{figure}

\newsection{Appendix B: Contour lines in the Dual Rotating Metric}
To get an idea of how the dual spinning ring coordinates $(x,y)$ span the space it is possible to obtain the flat space metric by taking (\ref{neutralring}) and setting $\lambda=0$. Having done this, the ring metric becomes
\begin{eqnarray}
ds^2=-\ud t^2+\frac{k^2}{(x-y)^2(1-\nu)}\left[\frac{(1+\nu)(1+\nu x^2y^2)}{(1-x^2)(1+\nu x^2)}\ud x^2+(1-x^2)(1+\nu y^2)\ud\phi^2\right. \nonumber \\
 \left. -\frac{(1+\nu)(1+\nu x^2y^2)}{(1-y^2)(1+\nu y^2)}\ud y^2- (1-y^2)(1+\nu x^2)\ud\psi^2\right]
\end{eqnarray}
Comparing this with the five dimensional flat space metric in spherical polar coordinates
\begin{equation}
ds^2=-dt^2+\ud r_1^2+r_1^2\ud\phi^2+\ud r_2^2+r_2^2\ud\psi^2
\end{equation}
allows the transformations between $r_1$ and $r_2$ and $x$ and $y$ to be determined. These are
\begin{equation}
r_1=\frac{k\sqrt{(1-x^2)(1+\nu y^2)}}{(x-y)\sqrt{1-\nu}} \hspace{3cm} r_2=\frac{k\sqrt{(y^2-1)(1+\nu x^2)}}{(x-y)\sqrt{1-\nu}}
\end{equation}
Theoretically these coordinate transformations can be used directly to plot lines of constant $x$ and $y$. In practice it is more useful to transform coordinates once again to get $r$ and $\theta$ in terms of $x$ and $y$. The $r_1,r_2$ coordinates are related to the $r,\theta$ coordinates by
\begin{eqnarray}
&& r^2 = r_1^2+r_2^2 = -\frac{k^2(1+\nu)(x+y)}{(x-y)(1-\nu)} \\
&& \tan^2{\theta} = \frac{r_2}{r_1} = \frac{(y^2-1)(1+\nu x^2)}{(1-x^2)(1+\nu y^2)}
\end{eqnarray}
Having obtained these coordinate transformations it is now possible to plot the contour lines for constant $x$ and $y$. A sample plot for $\nu=\frac{1}{2}$ is given in figure \ref{fig:contours}.

Unlike in the singly rotating ring, the contour lines of constant $y$ are now elliptical, rather than circular. This is due to the factors of $\nu$ that appear in the coordinate transformations. If $\nu$ is set to zero then the ellipses become circles, as one would expect since the $\nu\rightarrow 0$ limit of (\ref{neutralring}) reduces to the singly spinning ring metric. As $\nu$ is increased from zero, the ellipses become more elongated and the contours of constant $y$ become wider spaced.

\providecommand{\href}[2]{#2}\begingroup\raggedright\endgroup

\end{document}